\documentclass[prl,twocolumn]{revtex4}
\usepackage{bm}

\def\half	{\textstyle {1\over2}}
\def\phard	{p_{\rm hard}}

\def\Qs		{Q_s}
\def\tr		{{\rm tr}}

\def\k		{{\bf k}}
\def\n		{{\bf n}}
\def\p		{{\bf p}}
\def\v		{{\bf v}}
\def\x		{{\bf x}}

\def\A		{{\bf A}}
\def\E		{{\bf E}}
\def\B		{{\bf B}}

\begin{document}
\title{Apparent Thermalization due to Plasma Instabilities in
    Quark-Gluon Plasma}
\author{Peter Arnold and Jonathan Lenaghan}
\affiliation
    {%
    Department of Physics,
    University of Virginia,
    Charlottesville, Virginia 22901, USA
    }%
\author{Guy D. Moore}
\affiliation
    {%
    Department of Physics,
    McGill University, 3600 University St.,
    Montr\'eal QC H3A 2T8, Canada
    }%
\author{Laurence G. Yaffe}
\affiliation
    {%
    Department of Physics,
    University of Washington,
    Seattle, Washington 98195, USA
    }%

\date{\today}

\begin{abstract}
    Hydrodynamical modeling of heavy ion collisions at RHIC suggests that
    the quark-gluon plasma (QGP) ``thermalizes" in a remarkably short time
    scale, about 0.6 fm/c.
    We argue that this should be viewed as indicating fast isotropization,
    but not necessarily complete thermalization, of the non-equilibrium QGP.~
    Non-Abelian plasma instabilities can
    drive local isotropization of an anisotropic QGP on a time scale
    which is faster than ordinary perturbative scattering processes.
    As a result, we argue that theoretical expectations based on
    weak coupling analysis are not necessarily
    in conflict with hydrodynamic modeling of the early part
    of RHIC collisions,
    provided one recognizes the key role of non-Abelian plasma
    instabilities.
\end{abstract}


\maketitle


Hydrodynamic models of RHIC collisions (based on near-ideal fluids)
provide a good description of
a wide range of experimental data, including radial and elliptic flow
measurements,
provided one assumes that the initial partons thermalize in
about 0.6~fm/c \cite {RHIC-hydro}.
However, theoretical estimates based on perturbative
scattering processes yield expected thermalization times in the range
of 2.5 fm/c or above
\cite {thermalization times}.
What is the significance of this discrepancy?
Are weak-coupling analyses, which should be valid for asymptotically
high energy densities (and asymptotically large nuclei),
inapplicable at RHIC energies?
Perhaps so.
Or have dynamical processes which may be responsible for this fast
apparent thermalization not been correctly identified?
We will argue this is the case \cite {other efforts}.
Estimates
based on perturbative scattering
neglect essential dynamics:
the collective behavior associated with
non-Abelian plasma instabilities.
Such instabilities can produce large non-perturbative effects,
including apparent thermalization.
We discuss two qualitative lessons which emerge
from a weak-coupling analysis:
({\em i}) hydrodynamic behavior does {\it not} require
full thermalization --- isotropization of parton momenta in local fluid
rest frames suffices; and
({\em ii}) plasma instabilities can drive isotropization
at rates which are parametrically faster than 
perturbative scattering rates.

\vspace*{-5pt}
\section {Apparent Thermalization}

The thermalization time scale in a quark-gluon plasma,
defined as the inverse relaxation rate of arbitrarily small
departures from equilibrium,
depends on the rate of large-angle scattering
(and near-collinear splitting/join\-ing) processes
among quarks and gluons \cite {Arnold:2002zm}.
Parametrically, this time scale is \cite{foot:transport}
$\sim 26 \, [g^4 T \ln (2.4/g)]^{-1}$,
and for plausible values of RHIC parameters it is
hard to reconcile this time scale with the fast
apparent thermalization observed in RHIC collisions.
%
%
However, this time scale characterizing relaxation of asymptotically small
perturbations is {\em irrelevant} to the question of when
hydrodynamic models can be a good approximation
to the dynamics of a non-equilibrium quark-gluon plasma.
The essential assumption of ideal fluid hydrodynamic models
is that the stress tensor, in the local rest frame
at some point in the system, is nearly diagonal,%
\begin {equation}
    T_{ij} \approx p \> \delta_{ij} \,,
\label {eq:stress}
\end {equation}
with some equation of state relating the pressure $p$ to the
energy density.
But relation (\ref {eq:stress}) is just a statement of
{\em isotropy} (in the local fluid rest frame),
and is automatically true if typical excitations
have random directions --- even if their energy distribution
is far from thermal, or if the pressure $p$ differs from the equilibrium
pressure for a given energy density.
Consequently, understanding when a hydrodynamic model can first provide
a good approximation to the plasma dynamics is the same question as
understanding what dynamics drives isotropization.

\vspace*{-5pt}
\section{Plasma Instabilities}

To begin, we summarize
known results concerning gauge field instabilities
in anisotropic non-Abelian plasmas.
Further details may be found in
Refs.~\cite{Arnold:2002zm,Arnold:2003rq,instabilities,instabilities2}.

Let $\phard$ denote the characteristic momenta of typical excitations
in a non-equilibrium quark-gluon plasma.
(For example, in
the saturation scenario
\cite{saturation},
$\phard$ equals the saturation scale $\Qs$ at time $\Qs^{-1}$.)
We assume that $\phard$ is sufficiently large that these
excitations act like highly relativistic particles.
For time scales short compared to the mean free time between
large-angle scatterings of typical excitations
(and large compared to $\phard^{-1}$),
the natural framework for describing the dynamics
is collisionless
kinetic theory.  One splits the degrees of freedom into short wavelength
(or ``hard'' momentum) excitations which may be characterized
by a phase space distribution function $f(\p,\x,t)$,
and long wavelength (or ``soft'') gauge field modes which may be regarded
as forming a classical field.
For a non-Abelian theory, the resulting Boltzmann-Vlasov equation
has the form
\cite {Mrowczynski:1989np, foot:vlasov}
\begin {equation}
    (D_t + \v \cdot D_\x) \, f
    +
    \half g \, \{ (\E + \v \times \B)_i , \, \nabla_{p_i} f \}
    =
    0 \,.
\label {eq:BV}
\end {equation}
The corresponding Maxwell equations are
\begin {equation}
    (D_\nu \, F^{\mu\nu})_a = j^\mu_a
    \equiv
    g \, {\textstyle \int_\p} \> v^\mu \> \tr (t_a \, f) \,,
\label {eq:Max}
\end {equation}
with
$\int_\p \equiv \int {d^3\p \over (2\pi)^3}$,
$v^\mu \equiv (1,\hat \p)$,
and
$t_a$ a color generator.

Any distribution which is homogeneous (in space) and colorless,
combined with vanishing soft gauge field, gives a static solution
to Eqs.~(\ref {eq:BV},\ref {eq:Max}).
Perturbations about such solutions obey a linearized equation of motion
[obtained by linearizing Eq.~(\ref{eq:BV}) in deviations from the
static solution, solving for $\delta f$,
and plugging the result into Eq.~(\ref{eq:Max})]
which (after a space-time Fourier transform) has the form
\begin {equation}
    \left\{
	K^2 \, g^{\mu\nu} - K^\mu K^\nu + \Pi^{\mu\nu}(K)
    \right\}
    A_\nu(K) = 0 \,,
\label {eq:linear}
\end {equation}
where the wavevector $K^\mu \equiv (\omega,\k)$
\cite{foot:metric}.
The retarded gauge-field self-energy, generated by hard excitations, is
\begin {equation}
    \Pi^{\mu\nu}(K)
    =
    g^2 \int_\p {\partial f(\p) \over \partial p^l}
    \left[
	-v^\mu g^{l\nu} + {v^\mu v^\nu K^l \over v \cdot K - i\epsilon}
    \right] .
\label {eq:Pi}
\end {equation}
The zero-frequency spatial self-energy $\Pi_{ij}(0,\hat\k)$ depends on
the direction but not the magnitude of the spatial wavevector $\k$.
If $f(\p)$ is anisotropic but parity invariant,
then the self-energy matrix $\Pi(0,\hat\k)$
has a negative eigenvalue
for some directions of $\hat\k$.  This implies
that there are unstable solutions to the small fluctuation equation
(\ref {eq:linear}), {\it i.e.}, solutions
for which $\omega$ has a positive imaginary part
\cite{Arnold:2002zm,Arnold:2003rq}.
These are non-Abelian versions of Weibel instabilities
in ordinary plasma physics \cite{Weibel}.

Let $-\mu^2$ denote the most negative eigenvalue of $\Pi(0,\hat\k)$
(for any $\hat\k$).
Unstable modes have $|\k| < \mu$.
Let $\gamma$ denote the maximal growth rate of unstable modes.
If the hard particle distribution has $O(1)$ anisotropy \cite{foot:anisotropy}
then the maximum unstable wavevector $\mu$ and the maximum growth rate
$\gamma$ are both comparable to the effective mass $m_\infty$ of hard gluons,
\begin {equation}
 \mu^2 \sim \gamma^2 \sim m_\infty^2 = g^2 \int_\p \; {f(\p) \over |\p|} \,.
\label {eq:musq}
\end {equation}
If $\phard$ is the momentum scale which dominates the integral
(\ref {eq:musq}),
and $n \equiv \int_\p f(\p)$ is the spatial density of hard excitations,
then $m_\infty \sim g \, \sqrt {n/\phard}$.

To compare to perturbative scattering rates consider, for
example, a system with $n=O(\phard^3)$ --- the same parametric relation
as in equilibrium, where
$p \sim T$
and $n=O(T^3)$.
In this case
$m_\infty$, and hence
the instability growth rate $\gamma$ for $O(1)$ (or larger) anisotropy,
is $O(g\,\phard)$.
This rate is parametrically faster than the $O(g^4 \, \phard)$
rates for large-angle scattering or near-collinear splitting,
or even the $O(g^2 \, \phard)$ rate of small-angle scattering
\cite{Arnold:2002zm}.
More generally, for $O(1)$ anisotropy $\gamma$ is faster
than the large-angle scattering rate whenever
$n \ll \phard^3/g^2$ \cite{foot:compare}.
This inequality is satisfied parametrically
unless there is saturation, and even
in saturation scenarios, it is satisfied for
$t \gg \Qs^{-1}$ \cite{saturation}.

Numerical values depend, of course,
on the specific form of the anisotropic phase space distribution.
A simple example \cite{foot:example} involving
a typical particle energy of 1 GeV,
plasma energy density of 27 GeV/fm${}^3$,
a phase space distribution proportional to $(\p\cdot\hat {\bf z})^4$,
and $\alpha_s=0.5$ yields
$m_\infty \simeq 740$ MeV, and
$\gamma \simeq 280 \; \rm MeV = (0.7\; fm/c)^{-1}$
for $k \simeq 575$ MeV.
With more extreme anisotropy,
the growth rate $\gamma$ can approach $m_\infty$ itself
\cite{Arnold:2003rq}.
Yet other angular distributions can give slower growth rates.


Instabilities
will grow exponentially until some dynamics comes into play
which causes the amplitudes of unstable modes to saturate.
There are two natural possibilities for when this might happen
\cite{foot:expansion}.
If the unstable modes with wavenumbers of order $\mu$ grow until
the soft gauge field has an $O(\mu/g)$ amplitude
[or the field strength is $O(\mu^2/g)$], then
non-Abelian corrections to the linearized equation of motion
(\ref {eq:linear}) will become important and could substantially
affect the further evolution \cite{foot:gauge}.
In particular, one might expect these non-linearities to lead
to efficient transfer of energy from the unstable modes to
stable modes (with comparable wavenumber).

Alternatively, if instabilities do not saturate at $O(\mu/g)$
amplitudes, then they may continue growing until their amplitudes reach
the scale $\phard/g$
[and field strengths are $O(\mu \, \phard/g)$].
This is the point where the soft gauge field no longer acts
as a small perturbation on the motion of hard excitations.
To see this,
note that for this amplitude,
the gauge field part of a covariant derivative is just as large as
the ordinary derivative when acting on fluctuations with $O(\phard)$ momenta.
This is also the point where the energy density in the soft gauge field
becomes an $O(1)$ fraction of the total energy density,
$
    (F_{\rm soft}^{\mu\nu})^2
    \sim (\mu \, \phard/g)^2
    \sim n \, \phard
$.

There are reasons to believe the second alternative,
not the first, is correct.
The generalization to anisotropic plasmas of the HTL (``hard thermal loop'')
effective action is \cite{Mrowczynski:2004kv,foot:validity}
\begin {eqnarray}
\label {eq:S-HTL}
    S_{\rm eff}
    &=&
    -\int d^4x \>
    \biggl[
	{\textstyle {1\over4}}
	F^a_{\mu\nu} F^{a \mu\nu}
\\ && \qquad {}
	+
	g^2 \int_\p
	{f(\p) \over |\p|} \>
	F^a_{\alpha\mu}
	\left({v^\mu v^\nu \over (v\cdot D)^2}\right)_{ab}
	F_\nu^{b\alpha}
    \biggr] \,.
\nonumber
\end {eqnarray}
Evaluating this, explicitly, for arbitrary static fields
in order to
examine the corresponding effective potential
is not feasible.
But in the special case of fields which vary in only one spatial direction,
the effective action reduces to a simple local form.
Let $\hat\n$ denote the direction of the wavevector of the most
unstable mode.
For gauge fields which depend only on $\hat\n \cdot \x$,
one finds that the effective potential is
\cite{A&L,Iancu}
\begin {equation}
    V[\A(\hat\n \cdot \x)] = \int d^3x \>
    \left[
	{\textstyle {1\over4}} F^a_{ij} F^a_{ij}
	+
	\half A^a_i \, \Pi_{ij}(0,\hat \n) \, A^a_j
    \right] .
\end {equation}
When $\Pi(0,\hat\n)$ has a negative eigenvalue
this potential is unbounded below.
The runaway directions of steepest-descent correspond to Abelian field
configurations where the commutator terms in the field strength
$F_{ij}^a$ vanish.
This suggests that non-Abelian non-linearities may {\em not}
cause growing instabilities to saturate at the scale $\mu/g$,
provided the field configuration evolves toward an effectively
Abelian form which can continue rolling down the potential energy landscape.
This behavior has been seen in time-dependent numerical simulations
in 1+1 dimensions \cite{A&L,Rebhan:2004ur} ---
the instability locally ``Abelianizes'' and continues growing.
It is important to perform
full 3+1 dimensional simulations of the collisionless kinetic theory
(\ref{eq:BV},\ref{eq:Max})
to verify this conclusion.
Such simulations are in progress \cite{inprogress}.
Here, we shall assume that growth of instabilities, beyond the soft scale
$\mu/g$, will be confirmed.

\vspace*{-5pt}
\section{Isotropization}

Growing instabilities imply that the stress tensor
of the non-equilibrium system will receive
growing contributions from the soft gauge field.
The fastest growing linearized modes tend to
decrease the anisotropy in the total stress tensor \cite{instabilities}.
For example,
if the anisotropic hard particle distribution has a prolate form,
so that $T^{\rm hard}_{zz} \gg T^{\rm hard}_{xx}, \, T^{\rm hard}_{yy}$,
then the wavevectors of the fastest growing unstable modes
lie in the equatorial plane and the growth of these modes
produces a soft gauge field contribution to the stress tensor
which is oblate,
$T^{\rm soft}_{xx} \sim T^{\rm soft}_{yy} \gg T^{\rm soft}_{zz}$.
Conversely, for an oblate hard particle distribution,
the fastest growing unstable mode has its wavevector along the normal
direction and generates a prolate contribution to the stress.
Hence, even in the linearized regime, one can see that soft gauge
field instabilities push the system toward greater isotropy.
However, the soft contribution to the stress tensor
is small compared to the hard particle contribution,
and the back-reaction of the soft gauge field on the hard particles is
a tiny perturbation,
as long as the soft gauge field amplitude is much less than $O(\phard/g)$.

But if the soft gauge field amplitude reaches the scale $\phard/g$
then it no longer acts as a small perturbation to the dynamics
of hard excitations.
Recall that
the radius of curvature of an
excitation of momentum $p$ and charge $g$
in a magnetic field $B$ is $R = p/(gB)$.
If the radius of curvature is comparable to the magnetic field
coherence length $\mu^{-1}$,
which means $B \sim \mu p/g$,
then excitations of momentum $p$ will undergo $O(1)$ changes in direction
during traversals of any single coherence-length sized
magnetic field ``patch'' \cite{foot:patch}.

Therefore, if unstable soft gauge field modes with $O(\mu)$ wavevectors
grow until the field strength is $O(\mu \, \phard/g)$,
then typical excitations will experience $O(1)$ changes in direction
in times of order $\mu^{-1}$.
Excitations with differing momenta or colors will receive different
deflections from a given patch of (non-Abelian) magnetic field.
Excitations traversing different patches of magnetic field
(separated by $O(\mu^{-1})$) will receive nearly uncorrelated
deflections.

The net effect is that a soft gauge field with a non-perturbative amplitude
of order $\phard/g$ can effectively drive isotropization in the distribution
of typical hard excitations on a time scale which equals the coherence length
$\mu^{-1}$ of the soft gauge field.
And isotropization of the hard particle distribution will turn off
further growth in the soft gauge field
(since gauge field instabilities are absent for isotropic distributions).

As with all instabilities,
the time, or number of e-foldings, required for the soft gauge
field to become
large
depends on the size of initial ``seed'' amplitudes in the
relevant unstable modes.
The amplitude of the soft ($k \sim \mu$)
gauge field generated by a random color
charge distribution of the hard particles
can be estimated as
$A^2 \sim $ $g^2 n/\mu \sim g\sqrt{n \, \phard}$.
This is the smallest the seed field could be.
For densities from $n=O(\phard^3)$ 
up to the density limit $n=O(\phard^3/g^2)$ imposed by saturation,
$A \gtrsim O(g^{1/2} \phard)$.
This is only a factor of $g^{3/2}$ smaller than the nonperturbative
$O(\phard/g)$ amplitude.
Therefore, the number of e-foldings required for instabilities
to grow to this non-perturbative size is only of order $\ln (1/g)$.

Treating logs of $g$, for simplicity, as $O(1)$, this means that
if the initial anisotropy is $O(1)$ then
the characteristic growth time needed for unstable modes of the soft
gauge field to reach the non-perturbative amplitude $\phard/g$ is only
of order $\gamma^{-1}$.
The resulting soft gauge field then drives isotropization
of the hard particle distribution on a comparable $\gamma^{-1}$ time scale.
Therefore, (up to logs of $g$ and factors of order one),
the time scale for isotropization of the hard particle distribution
is the {\em same} as the (inverse) instability growth rate $\gamma^{-1}$
\cite{foot:violent}.

In numerical simulations of ordinary non-relativistic plasmas,
essentially the same process of instability-driven isotropization
has been observed \cite{Califano},
with the growth of magnetic instabilities driving large reductions
in anisotropy once the magnetic fields reach critical strength.
(These simulations allowed three-dimensional momentum space variations,
but assumed translation invariance in one spatial direction.)
Various QGP numerical simulations
\cite{QGP numerics}
have failed to see any sign of this instability-driven dynamics
because they did not allow full three-dimensional variations.

Although we have focused on the ability of non-perturbative soft
gauge fields to generate large changes in directions of hard excitations,
it should be noted that $\mu^{-1}$ is also the characteristic time
scale for $O(1)$ changes in energies of hard excitations.
This is inevitable, given the fact, noted earlier, that
when the soft gauge field reaches the non-perturbative amplitude $\phard/g$
its energy density is comparable to the energy density in the hard
excitations.
But it may also be seen directly by noting that
chromoelectric fields generated during the growth of instabilities
will be comparable in size to chromomagnetic fields
(since the growth rate of unstable modes is comparable to their wavenumbers
for $O(1)$ anisotropy).
So chromoelectric fields will reach the same
$O(\mu \, \phard/g)$ size as magnetic fields
--- which means that an excitation traveling
a distance $\mu^{-1}$ will have work of order $\phard$ done on it by the
soft gauge field.
Of course, this time scale for $O(1)$ changes in energy may be very
different (and much shorter) than the time scale for true thermalization,
as defined by a near-thermal energy distribution of excitations over
a parametrically large dynamic range.

\vspace*{-5pt}
\section{Conclusions}

We have argued that ``early thermalization'' in heavy-ion
collisions is more properly interpreted as evidence of fast isotropization
in the distribution of excitations.
And we have argued that non-Abelian plasma instabilities can drive
isotropization at a rate which is parametrically fast
compared to perturbative scattering rates.
Consequently, we see no reason to view the fast onset of
hydrodynamic behavior in RHIC collisions as necessarily in conflict with
theoretical expectations based on weak-coupling analysis of a
quark-gluon plasma, provided one properly accounts for the
effects of non-perturbative plasma instabilities.
Further study of the scenario we have sketched is certainly needed;
in particular full three-dimensional non-Abelian Boltzmann-Vlasov simulations
with appropriate initial conditions should be conducted.


\begin{thebibliography}{}
\bibitem{foot:compare}
The large-angle scattering rate is $n\sigma (1{+}f)$, where
$\sigma \sim g^4/\phard^2$, and $1{+}f$ is a Bose enhancement
factor, with $f \sim n/\phard^3$ for $O(1)$ anisotropy.
Comparison to $m_\infty$ gives the stated inequality.
\bibitem{foot:gauge}
    Estimates of gauge field amplitudes should be understood
    as applying in gauges, such as Coulomb gauge, which make
    the gauge field as smooth as possible.
\bibitem{foot:validity}
The derivation of the effective action (\ref {eq:S-HTL})
    assumes that the gauge field is perturbative on the scale
    of the hard excitations.
    It ceases to be valid if the gauge field grows to a non-perturbative
    amplitude of order $\phard/g$.
\bibitem{foot:patch}
     If the coherence length $\mu^{-1}$ were less
     than the radius of curvature, then a hard excitation
     crossing any
     particular coherence-length sized patch of magnetic field
     will receive a small $O(gB/\mu p)$ deflection.
     Deflections from different patches will add randomly,
     so traversal of
     $O[(\mu p/gB)^2]\gg 1$ patches will be required to
     receive an $O(1)$ change in direction.
     This is the usual case where the soft gauge field is
     perturbatively small on the scale of hard momenta.
\end{thebibliography}
\end {document}